\documentstyle[12pt]{article}

\begin{document}

\topmargin -8mm
\oddsidemargin 0mm

\setcounter{page}{1}
\vspace{1cm}
\begin{center}
\bf PRESENT SITUATION OF DIFFRACTED X-RAY RADIATION AND RESONANCE 
(COHERENT) TRANSITION RADIATION INDUCED BY HIGH ENERGY CHARGED PARTICLES 
IN FREQUENCIES REGION EXIDING ATOMIC ONE.\\
\vspace{3mm}
{\bf M.L. Ter-Mikayelyan}\\
\vspace{3mm}
{\small Institute for Physical Research of Armenian Academy of Science
378410, Ashtarak-2, Armenia.}\\
\end{center}
\vspace{3mm}
\indent
\begin{abstract}
\indent
~~~~The review is devoted to the modern investigations of 
electromagnetic radiation by relativistic charged particles 
propagating with constant velocity through the periodic media. 
Two examples of radiation are considered in this review, the first 
corresponds to the Bragg scattering of charged particles pseudophotons 
in crystals, the second one to the Fresnel scattering of pseudophotons 
in periodic medium. Both examples play essential role in construction new 
compact tunable sources in X-ray region.
\end{abstract}

\indent
{\bf 1. Introduction}\\
\indent
High-energy electromagnetic processes in medium have been summarized
in my monograph in 1969 year and translated into English in 1972 [1]. 
(The numbering of chapters, paragraphs, formulas and total text of
English edition are the same as in Russian version.) Since that time 
this field of high-energy physics developed catastrophically fast in 
many various directions, including quantum chromodynanic (see [2] and 
references therein). During that period of time many original 
publications, review articles and books have been published. 
The number of original publications is impossible to estimate. 
Only the physicists from former Soviet Union published in Physics 
Usphekhi since 1972 ten reviews and eleven monographs in various 
publishing houses. Numerous conferences, workshops and seminars in 
different countries were devoted to these problems. For example: 
traditional conferences in Russia (Tomsk, RREPS1993-1995-1997-1999), 
(see [3] and references therein), in Germany [4] (Tabarz, 1998) etc.

\indent
All these problems of high-energy radiation physics are based on the 
following underlying idea; the length of trajectory (coherence length) 
along the trajectory of initiating reaction particle increases with energy 
of incoming particle and the directionality of process (see Appendix). 
The history of this concept has been published in Russian by E.L. Feinberg. 
in his well known paper in Priroda  [5], (see translation of [5] into 
English in Appendix). Among numerous problems from this field I will 
review in this paper only two; the diffracted X radiation (DXR) 
and the resonance transition radiation (RTR).

\vspace{3mm}
\indent
{\bf 2. Diffracted X Radiation (DXR)}\\
\indent
DXR has been considered in 1969 and included in Chapt.5 of 
book [1]. The theoretical consideration can be very simplified, 
if we follow the corresponding theory of X-ray scattering in crystals.  
Expanding electrical field of fast moving particle in time and variable 
part of dielectric susceptibility $\varepsilon_{1}(r)$ over vectors 
$\vec {g}$ of the reciprocal lattice.
\begin{equation}
\label{AA}
\varepsilon=\varepsilon_{0}+\varepsilon _{1}(\vec{r});~~
\varepsilon _{1}(\vec{r})=\sum n_{\vec{b}} \exp(i\vec{g}\vec{r})
\end{equation}
and using the theory of  X ray scattering we get the following expression 
for scattered electric field at a distance $R_{0}$
\begin{equation}
\label{AB}
\vec{E^{'}_{\omega}}(\vec{R_{0}})=\frac {e}{\varepsilon_{0}R_{0}}
\sum_{\vec{b}} n_{\vec{b}}
\left[\vec{k^{'}}\times\vec{k^{'}}\times\left(\frac{\omega \vec{v}}{c^{2}}+
\frac{\vec{g}}{\varepsilon}\right)\right]\cdot
\frac{\delta [\omega-(\vec{k^{'}}-\vec{g})\vec{v}]}
{(\vec{k^{'}}-\vec{g})^{2}-(\omega^{2}/c^{2})\varepsilon_{0}}.
\end{equation}
Polarization of D.X.R. is linear and is given by expression (2). 
The radiation angle is determined, assuming the argument of the 
$\delta$-function in (2) to be equal to zero
\begin{equation}
\label{AC}
\cos \theta=\frac{c}{v\sqrt{\varepsilon_{0}}}+
\frac{(\vec{g}\vec{v})c}{\omega v \sqrt{\varepsilon_{0}}}
\end{equation}
where $\theta$ is the angle between velocity and scattered photon. 
Expression (3) often used in practice, for measuring the dependence of 
emitted photon energy upon the scattering angle $\theta$. The equality (3) 
follows from energy-momentum conservation laws for photon radiation in 
a crystal if we take into account, that crystal can receive momentum 
inverse proportional to the length of periodicity.

\indent
Intensity is given by following expressions
\begin{equation}
\label{AD}
d\vec{I}_{\omega , \vec{n}}=c\sqrt{\varepsilon_{0}}
|\vec{E}^{'}_{\omega}|^{2}R^{2}_{0}d\omega d\Omega
\end{equation}
\begin{equation}
\label{AE}
d\vec{I}_{\omega , \vec{n}}=\frac{e^{2}\omega^{2}T}
{2\pi \varepsilon^{5/2}c}
\sum_{\vec{b}} n_{\vec{b}}^{2}\left|\vec{k^{'}}
\times\left(\frac{\omega \varepsilon}{c^{2}}\vec{v}+
\vec{g}\right)\right|^{2}\times
\frac{\delta [\omega-(\vec{k^{'}}-\vec{g})\vec{v}]}
{[(\vec{k^{'}}-\vec{g})^{2}-(\varepsilon_{0}/c^{2})\omega^{2}]^{2}}
d\omega d\Omega ~.
\end{equation}
Formulae for polarization (2), angular distribution (3), and the 
radiated energy in frequency interval $d\omega$ (5) are the base of 
kinematical theory [1] and have been confirmed in numerous theoretical 
and experimental papers.

\indent 
The first experimental investigations of DXR have been done in Tomsk, 
by group of A.P. Potylitsyn [6,7], the second in Yerevan [8], third in 
Kharkov [9]. DXR initiated by protons has been observed in [10]. 
In papers [11,12,13] was shown that kinematical theory [1] may be 
sufficient to explain modern experimental results on spectral and 
angular distributions of DXR, as well as absolute differential yield. In papers [14] 
quantum theory of DXR has been developed and has been shown, that for 
relativistic particles, if recoil due to the photon emission is small, 
the quantum expressions coincides with corresponding expressions presented 
in [1]. I will mention that like dynamical theory of X-ray scattering, 
the corresponding theory in DXR has been developed in series of publications 
and included in monographs [15,16]. This type of radiation is referred in 
literature under names quasi-cherencov, parametric and even polarization 
radiation. To simplify the terminology I will use for this type of radiation 
the name dynamical theory of DXR The name PXR often used in literature 
for both type of radiation is meaningless and doesn't correspond to the
nature of radiation (I thank H.Nitta for this comment).
During the last years new experimental investigation has been accomplished, 
which improved our knowledge of DXR. The remarkable features of DXR 
(monochromatic with continuously variable wavelength, propagation direction 
well separated from the electrons beam one, small energy and angular spread 
of the order of magnitude one over gamma, and at last the small sizes of setup) 
suggest that DXR can be an perspective X- ray source in future. A series of paper 
[17,18] has been published recently by joint group from Institute 
fuer Kernphysik in Darmstadt, from Kharkov, Rossendorf and Johannesburg, 
using superconducting linear accelerator S-DALLINAC with electron energy 
below 10 MeV. The line width of 8.98 keV DXR for electron energy 6 Mev 
has been measured applying an absorption technique using a copper foil 
and tuning the energy of the DXR peak across the K-absorption edge of copper. 
The spectral density in the peak deduced from experiment was,
$I=0.95\times 10^{-7}$ photons/(electron sr.eV), and linewidth 48 eV (see Fig.1). 
In paper [19] upper limits of line width of DXR 1.2 eV and 3.5 eV have been 
determined for (111) and (022) reflections of silicon at photon energies 
of 4966 eV and 8332 eV. Investigations of the line width of DXR at the Mainz 
Microtron MMI a relative energy width ($\Delta E/E=10^{-5}$ 
should be reached for the 
silicon (333) reflection [19,20]. The spectral and angular distribution of 
DXR have been studied mostly in silicon and diamond crystals over a 
range from a few MeV up to several GeV of electrons and are in
consistent with the theory [1].

\indent
The last experiments carried out by physicists from Germany 
(Werner Heisenrberg Institute and Institute fuer Kernphysik), 
[21,22] has shown close to $100\%$ linearly polarization at every 
single point of photons angular distribution, with agreement 
with the theory [1]. In these investigations polarization has been 
analyzed by means of novel method of polarimetry exploiting directional 
information of the photoeffect in a charge coupled device consisting of
$1.3\times 10^{6}$ square pixels of 6.8mkm [23]. 
The advent of such devices opens 
a promising route towards a universal X-ray detector for simultaneous 
imaging, spectroscopy and polarimetry.
The angular distributions of DXR polarization directions, calculated 
recently [24] on the base of the theory [1], are close to the 
experimental and calculated data presented in [21,22] for DXR 
in forward and backward hemispheres, but in disagreement with calculations 
[21,22] for DXR polarization at  right angle. The disagreement between 
calculations is due because in [21,22] the longitudinal density effect 
[25] has been neglected (private remark from A.V. Schagin). 
The discrepancy of both calculations with experiment [26] for 
polarization in forward hemisphere emission remains unsolved yet and 
new measurement is needed (private communication from R. Kottaus).

\indent
The physicist from Tomsk investigated influences of temperature on DXR 
intriducing Debay-Waller factor in expression (5). They obtained good 
agreement with experimental data [31]. In first publication devoted to 
the influence of acoustic waves and gradient of temperature on DXR shows 
that the intensity of DXR may be increased several times [32] (see Fig.2). 
Nevertheless intensity of DXR attained in laboratories in several keV 
domains is the same order of magnitude as synchrotron radiation of big 
accelerarators (see [3,4,33] and references therein).

\indent
Concluding this short review of DXR I will notice that more 
complicated theoretical and experimental problems remains unsolved. 
In particular the region of applicability of dynamical DXR theory 
and its correspondence with experiments [27,28,29] has not been investigated 
seriously. During the International Workshop on Radiation Physics, in 
Tabartz [4] Prof. Baryshevsky V. affirmed that dynamical version of DXR 
is necessary to understand experimental data [29]. On the other hand Prof. 
N. Nasonov  maintains opposite statement.I. Feranchuk  and A. Ivashin  
incorporated quantitatively electron multiple scattering and photon 
absorption in kinematics theory [30], but more subtle theoretical 
treatment is necessary.

\vspace{12mm}
\indent
{\bf 3. Resonance Transition Radiation (RTR)}\\
\indent
The well known expression for Transition Radiation (TR) introduced 
in physics in 1946 by V. Ginzburg and I. Frank received a new development,
when it was investigated in radiation frequencies exceeding optical 
[34,35]. At that time the longitudinal density effect [25] and coherence
length concept introduced in high-energy radiation processes in papers 
[36] were well known and the results of papers [34,35] can be easily 
understood [37] and derived from [25]. The problems arise when the 
expression for transition radiation, which is valid only for one 
interface, tried applying for many periodically spaced interfaces 
and in limiting case for periodic medium. It must be taken into account, 
that nonrelativistic charged particles propagating through periodic 
medium will emit photons with frequencies proportional to the frequency 
of propagation the periodicity of medium (resonance condition). 
For relativistic particles because of Lorenz transformation we get 
the following resonance condition
\begin{equation}
\label{AF}
\cos\theta=\frac{c}{v\sqrt{\varepsilon_{0}}}-
\frac{2\pi r c }{l\omega\sqrt{\varepsilon_{0}}}\cos\psi
\end{equation}
where ($\omega$-frequency of radiation, v -velocity of particle, 
($\varepsilon_{0}$ - effective dielectric susceptibility, ($\theta$
 - angle between incoming 
particle velocity and direction of radiation, ($\psi$ - angle of incidence 
of charged particle onto the one-dimensional periodic medium, $l$-period 
of medium, $r$ - number of emitting harmonic. The resonance condition (6) 
can be easily derived using the energy-momentum conservation laws in 
periodic medium. This kind of TR was termed as RTR. For $l\rightarrow 
\propto$ we get the
well-known Tamm-Frank expression for Vavilov-Cherenkov radiation in 
homogeneous medium. RTR consist of overlapping radiated harmonics, 
each has its threshold in energy of radiating particle and depends 
on parameters of medium. Theory of RTR has been published in my 
article presented for publication by L. Landau to Dokladi Acad. Nauk 
in 1960 [38] and published in Nuclear Physics in 1961 [39]. For 
periodic medium radiation on $r$-harmonic appears when the particle 
velocity exceeds the group velocity of corresponding photons.\footnote
{Beginning from I. Frank proposal, many physicists tried to increase the
TR intensity from one interface using many foils. The calculations
in optical region were cumbersome and negaive, because of neglecting
resonance condition. This problem was similar to the corresponding
one in saturation problem of ionization losses solved by E. Fermi, 
who enlarged the I. Tamm calculation for Cherenkov-Vavilov radiation.}

\indent 
~~~~~The most convenient medium for experimental investigation is laminar 
periodic medium with many plates (Fig.3). For laminar medium consisted 
with two different plats $\sqrt\varepsilon_{0}$ has simple form
\begin{equation}
\label{AG}
\sqrt\varepsilon_{0}=\frac{l_{1}\sqrt{\varepsilon_{1}}+
l_{2}\sqrt{\varepsilon_{2}}}{l}
\end{equation} 
where $l_{1}$ and $l_{2}$  are thickness, $l=l_{1}+l_{2}$ and $\varepsilon_{1}$
and $\varepsilon_{2}$  are dielectric susceptibilities 
of plates. For frequencies much more higher optical frequencies from 
inequality $|\cos\theta|\le 1$, for each harmonic we get
\begin{equation}
\label{BA}
\omega_{max}=\frac{4\pi cr}{l}\left(\frac{E}{mc^{2}}\right)^{2}
\ge\omega\ge\frac{l\omega^{2}_{0}}{4\pi cr}=\omega_{min} ~,
\end{equation}
where $\omega_{0}$ is the plasma frequency 
\begin{equation}
\label{BB}
\omega^{2}_{0}=\frac{4\pi NZe^{2}}{m_{e}}
\end{equation}
and for laminar medium
\begin{equation}
\label{BC}
NZ=(N_{1}Z_{1}l_{1}+N_{2}Z_{2}l_{2})/l
\end{equation}
From (8) we get the threshold energy for radiation harmonic of 
$r$-number. The intensity of RTR is given by following expression 
(see formula $(28.92^{*})$ from [1] or paper [39]).
\begin{eqnarray}
\label{BE}
dI_{\omega,\theta}&=&\frac{e^{2}\theta^{3}d\theta d\omega}{2\pi c}
\left|\frac{\varepsilon_{2}-\varepsilon_{1}}
{\left(1-\frac{v}{c}\sqrt{\varepsilon_{1}}\cos\theta\right)
\left(1-\frac{v}{c}\sqrt{\varepsilon_{2}}\cos\theta\right)}\right|^{2}
\times \nonumber\\
&&\times \sin^{2}\left[\frac{l_{1}\omega}{2c}
\left(1-\frac{v}{c}\sqrt{\varepsilon_{1}}\cos\theta\right)\right]
\frac{\sin^{2}\frac{n\beta}{2}}{\sin^{2}\beta}~.
\end{eqnarray}
where the first term corresponds to Ginzburg-Frank transition radiation, 
the second corresponds to the interference of radiation from two 
interfaces of  one plate and the last term corresponds to the coherent 
summation of radiation from n-plates. The quantity $\beta$ equals
\begin{equation}
\label{BF}
\beta=\left(1-\frac{v}{c}\sqrt{\varepsilon_{1}}\cos\theta\right)
\frac{\omega l_{1}}{2v}+
\left(1-\frac{v}{c}\sqrt{\varepsilon_{2}}\cos\theta\right)
\frac{\omega l_{2}}{2v}
\end{equation}

If the number of plates increases the last term can be substituted by
delta function and we get the resonance condition (6). The theory
of RTR depends dramatically on coherence length, if for example 
coherence length exceeds the distances between two interfaces in 
a plate the radiation from two interfaces must be summed coherently. 
The same result takes place for total periodic medium. Radiation from 
plates will sum independently if the distances between plates exceed 
the coherence length. We shall discuss the related experiments later. 
The RTR including absorption influence and multiple scattering effects
has been discussed in  [39,1]. But in that time (sixteenths years) 
we were interested to apply RTR for construction a new type of 
counters for very high energies of particles were Cherencov 
detectors were insensitive. These new detectors has been constructed 
by group of F.R. Arutunian in 1963 [40,41]. In following 
investigations the property of these detectors were improved and 
reviewed in many publications [42,43,44,45]. They are used now for 
identification of particles in modern high-energy accelerators (see 
for example [46] and references therein).

\indent
Since 1985 RTR received a new impact for developing in different domain 
of physics. Joint group of physicist from Stanford University and Livermore 
Lawrence Laboratory investigated RTR using the linear accelerator with 
energy of electrons equal to 17.2, 25, 54 MeV to produce photons in keV
region [47]. Stack of Be, C, Al foils consistent each from18 foils
with thickness 1 $\mu$m separated in distance 0.75mm (for carbon) and 1.5mm 
(Be and Al) have been used. Experimental data for RTR intensity angular 
and spectral distributions presented in [47] confirms the theory of 
interference at the interfaces of a single foil. The authors 
assert that an easy-to-tune source of intense polarized monochromatic 
radiation holds much promise for submicron lithography. For instance, 
a 0.5 $\mu$m resolution was reported in [48]. Though in [47-49] no 
interference effects were observed with radiation from different 
beryllium foils (see Fig.4), the same authors noticed an unusual 
interference pattern in [50]. Interesting observations of interference 
effects in RTR were made by French researchers [51] at the electron 
accelerator in Sacle (see Fig.5). The achievement of 0.3 $\mu$m resolution 
was reported in [52]. In the soft range of the spectrum (1-3 keV) the 
RTR spatial distribution for electron energy of 50-228 MeV was observed 
in [53]. Here attention is drawn to the fact that RTR results through 
the whole radiator stack, which forms the periodic structure, and 
concentrates in the solid cone whose angle grows with electron energy. 
Changes in the periodic structure parameters (e.g. electron energy, 
structure size) suppress interference effects and give rise to TR 
for which the emission angle, in coincidence with the TR theory, is
inversely proportional to the electron energy (see Fig.6). Teams 
from universities of Kyoto, Tohoku, Hyrosimo, Tokyo (I. Endo et al.) 
and Tomsk Institute for Nuclear Physics (A.P. Potylitsin et al.) in 
cooperation with various Japanese firms have made a lot of research 
into RTR at accelerators in Japan [54-58]. The goal of the research 
was not only to investigate resonance effects in RTR but also to 
determine the parameters of the electron beam and active medium best
suitable for practical implementations of RTR.
 
\vspace{3mm}
\indent
{\bf 4. The radiation of moving particles on complex structures (DXR+ RTR)}\\
\indent
The use of complex periodic structures was first suggested in the 90s 
when it became clear that the development of effective kiloelectron-volt 
generators requires increased intensities of DXR+ RTR [59]. Russian 
and Japanese physicists joined efforts to conduct research in this area. 
Papers [60] present the results of the irradiation of a target consisted 
of three crystals 16 (m thick with 800-MeV electrons (the synchrotron in 
Tomsk (Fig.7a)) and with 900-MeV electrons (the linear accelerator in Tokyo). 
Besides DXR, in the first crystal layer of the stack the electron beam 
generates RTR, which undergoes Bragg diffraction on the following crystal 
layers. This gives rise to the effective growth of emission (a 1.7-times 
increase was observed in the experiment). It should be noted that with 
the emission angle of the same order the RTR intensity and spectral 
width is much greater than that of DXR. The difference is that RTR 
follows the electron path, while DXR propagates along Bragg's angles 
of refraction from the corresponding crystal planes in the reciprocal 
lattice. The authors introduced a new name for this type of emission - 
parametric (diffracted) RTR (substituting letter P by D (DRTR)), 
which we will further keep to. The next experiment [61] dealt with a 
900-MeV electron beam and a target consisted of ten mylar foils and 
graphite crystal (Fig. 7b). The authors assert that even with a few 
foils DRTR follows Bragg's angles and is much more intensive than DXR. 
The last joint works of these authors [62,63] investigates radiation in the 
keV spectral range in a periodic medium consisting with crystalline plates. 
A 900 MeV electron beam and a target of 1 to 100 plates of monocrystal 
silicon were used in the experiment (see Fig.7c). The DRTR intensity 
of 35.5 keV photons proved to be comparable to that of synchrotron emission 
caused by a 1.7 GeV electron beam. The paper also considers the relationship 
between the radiation intensity and the number of plates - the issue that was 
discussed earlier in [57]. Papers [62], [63] and [4] (the last citation 
refers related papers presented at the meeting in Tabarz, Germany, 1998) 
build a theory that establishes a link between RTR and diffraction radiation 
that caused by a charged particle flying over a surface with periodic 
irregularities. In the experiment an electron beam propagated over a 
GaAs plate whose surface had 300 identical strips which were 10 $\mu m$ wide, 
100 $\mu m$ high and spaced 33 $\mu m$ apart. The authors observed radiation that 
consisted of DXR and DRTR, the intensity of the latter being much higher.

\indent
A great deal of theoretical papers discussing interference of various 
kinds of radiation has been published recently. Paper [64] offers a 
method of separating DXR and DRTR. Paper [65] shows that DXR 
output in mosaic crystal is the same as in a perfect crystal, 
and DRTR output is much higher. Diffraction of TR on a crystal 
structure is considered theoretically in [66]. Several relevant theoretical 
papers were also presented at the recent international conferences [3,4].
 
\vspace{3mm}
\indent
{\bf 5. Conclusion}\\
\indent
As I have already noted in Introduction, I have elucidated, out of a 
large set of questions, only DXR and RTR, which are developed recently 
in the numerous physical community. I hope to focus on other problems 
of yigh energy electromagnetic processes in medium in my forthcoming 
reviews. Aouthor will be very grateful for any comments and suggestions 
to improve this review. 
 
\vspace{3mm}
\indent
{\bf 6. Acknowledgments}\\
\indent
I am very grateful to Organizing Committee of RREPS'99 for the invitation 
to take part in the traditional symposium on the lake Baikal organized by 
Tomsk Nuclear Physics Institute. I would like to express a special 
gratitude to A.P. Potylitsyn, Yu. L. Pivovarov and L. Puzyrevich whose 
hospitality is very hard to overestimate. The work was performed within 
the program of the Ministry of Education and Science of the Republic of 
Armenia, grant $\#96-772$ and the INTAS grant $\#99-392$.

\newpage
{\bf APPENDIX}
\begin{center}
\bf EFFECT CONFIRMED 40 YEARS LATER.\\
\vspace{3mm}
{\bf "Nature", Russian Academy of Sciences, 1994, N1, pp. 30-33}\\
{\bf E.L. Feinberg}\\
\vspace{3mm}
{\bf Corresponding member of RAS, P.N. Lebedev Physical 
Institute of RAS}\\
\end{center}
\vspace{3mm}
It was recently reported that in the accelerator center of Stanford 
University (SLAC) direct experimental evidence was obtained of the 
suppression of bremsstrahlung of relativistic particles in an  medium [
1], theoretically developed by L.D. Landau and I.Ya. Pomeranchuk 40 
years ago [2]. This experiment confirms already the third of the 
important effects predicted in a series of works of Soviet theoreticians 
in 1952-1954. All these effects are bound by a common physical idea 
(or a basis), although they are displayed in different interactions of 
high-energy particles, and not only electromagnetic, but, as well, 
nuclear. This basis was built in 1952 in the Ph.D-thesis of M.L. 
Ter-Mikayelian [3], the post-graduate, of that time, in the Theoretical 
Department of the P.N. Lebedev Physical Institute, AS of USSR. The work 
was devoted to the investigation of the bremsstrahlung but not on single 
atom, as was studied before; it was considered in a medium, specifically, 
in a crystal.

\indent 
The result of this work which seemed at that time paraamorphousdoxical, 
consisted in a statement that at very high energies, when the wavelength 
of either the emitted photon or the electron is tens of millions, 
milliards times shorter than the mean interatomic distance in the medium, 
the usual radiation pattern changes dramatically. Particularly, if the 
motion occurs along the crystal axis, this radiation may many times exceed 
on individual atoms. The process in this case is widely extended in space 
and includes a domain with characteristic sizes many times exceeding the 
interatomic separations. All atoms of the crystal in this domain act 
coherently, and as a result, the radiation is enhanced significantly. 
This length was, naturally, termed as the coherence length.

\indent
The work under discussion had forerunners. A possibility of the influence
of crystalline structure on the bremsstrahlung of fast particles was
discussed by B. Ferretti in 1950, and still earlier, in 1935, by
E.Williams, who developed independently the well-known, in theoretical 
physics, Weizsecker-Willians method. However, either the work of
Ferretti or the notation of Williams (who obtained, by the way,
an incorrect sign of the effect) remained unpersuasive and did 
not attract the attention until Ter-Mikayelian succeeded to show 
that the bases of the process are paradoxical (for that time) physical 
causes which turned out to have much wider significance than the 
explanation of radiation in crystals. This lead to the development
of a new direction involving various processes in high-energy physics.
Now attempts are made to apply these results to processes of  of
high-energy hadrons inside the nucleus regarded as a material medium.

\vspace{3mm}
\indent
{\bf Coherence length}\\
\indent
Ter-Mikayelian succeeded to show [5] that in the processes at high 
energies where the particles are scattered at small angles 
(decreasing with the icrease of the energy), the longitudinal momentum, 
$q_{||}$, transferred to the target, drops, and, consequently, according to
the uncertainty relation, $\Delta q_{||}\Delta x_{||} \ge \hbar$
 , the longitudinal distance  $\Delta x_{||}$ involved in the 
process increases with the energy. Therefore, it is the coherence 
length, $L_{coh} \sim \hbar/q_{||}$ , rather than the wavelength 
of the particle, that can be a 
measure of the size of domain, relevant for the effect. When speaking
not especially of a crystal it would be reasonable to term this length
otherwise, namelythe zone or length of process formation. In 1952 this 
looked incredible and even absurd since it was accepted that characteristic 
distances of formation of electromagnetic processes are of the same order as 
the wavelengths of particles involved (or the atomic sizes).

\indent
In order to illustrate this nontrivial result let us consider the process 
of bremsstrahlung of a photon with energy  $\hbar\omega $ 
and momentum  $\hbar \vec{k}$ on a fixed 
coulombian center. Let $E_{1}$ and $\vec{p_{1}}$ be the 
initial energy and momentum of 
the radiating relativistic particle of mass {m} while $E_{2}$ 
and  $\vec{p_{1}}$ are the same 
quantities in its final state. Let us then use the energy 
and momentum conservation laws,\\
\rightline {$E_{1}-E_{2}=\hbar\omega
~~~~~~~~~~~~~~~~~~~~~~~~~~~~~~~~~~~~~~~~~~~~~~~~~~~~~~~~~(1)$}
\rightline {$\vec{p_{1}}-\vec{p_{2}}-\hbar \vec{k}=\vec{q}
~~~~~~~~~~~~~~~~~~~~~~~~~~~~~~~~~~~~~~~~~~~~~~~~~~~~~~~~~(2)$}
Where $\vec{q}$ is the momentum transferred to the nucleus, and project the 
latter on the initial direction of particle's motion. For this purpose we 
multiply Eq. (2) by the initial velocity of the particle, $\vec{v_{1}}$:\\
\indent
$ ~~~~~~~~~~~~~~~\vec{v_{1}}\delta \vec{p}-\hbar \vec{k} \vec{v_{1}}=
\vec{v_{1}}\vec{q}\approx |\vec{v_{1}}|q_{\|}~,$\\
Where $\delta\vec{p}=\vec{p_{1}}-\vec{p_{2}}~$, and  $q_{\|}$ 
is the longitudinal momentum transferred along the motion of
the emitting particle.

\indent
Since $\vec{v}\delta\vec{p}=\delta E=E_{1}-E_{2}=\hbar\omega$, 
we have for small energy variation, $\hbar\omega\ll E$, of the initially 
ultrarelativistic particle ( $|\vec{v_{1}}|\approx|\vec{v_{2}}|\approx c$ 
and ${k}=\omega/c$),\\
\rightline
{{\large$q_{\|}=\frac{\hbar\omega}{c}\left(1-\frac{v}{c}\cos\theta\right)$}.
~~~~~~~~~~~~~~~~~~~~~~~~~~~~~~~~~~~~~~~~~~~~~~~~~~~~~~~~~(3)}
Where $\theta$ is the angle between the emitted and the direction of 
motion, $ \vec {v_{1}}$~, 
of the emitting particle. As the radiation at high energies is known to 
be sharply directed, the obtained formula should be considered as small
$\theta$ . For $\theta\ll\sqrt{1-v^{2}/c^{2}}$ \\
\rightline
{{\large $q_{\|}=\frac{\hbar\omega}{c}\left(1-\frac{v}{c}\right)$}.
~~~~~~~~~~~~~~~~~~~~~~~~~~~~~~~~~~~~~~~~~~~~~~~~~~~~~~~~~(4)}
As it was mentioned above, the coherence length (formation zone)
along the path of the emitting particle amounts by the order of magnitude to\\ 
\rightline
{{\large$L_{coh}\approx \frac{\hbar}{q_{\|}}\approx\frac{c}
{\omega(1-v/c)}\approx\frac{E^{2}}{m^{2}c^{3}\omega}$}.
~~~~~~~~~~~~~~~~~~~~~~~~~~~~~~~~~~~~~~~~~~~~~~~~~~~~~~~~~(5)}
With the time of passing the zone being equal to\\ 
\rightline
{{\large$t_{coh}\approx\frac{L_{coh}}{v}\approx\frac{E^{2}}{m^{2}c^{4}\omega}$}.
~~~~~~~~~~~~~~~~~~~~~~~~~~~~~~~~~~~~~~~~~~~~~~~~~~~~~~~~~(6)}\\
\indent
This means that at $v\rightarrow c$ the quantity $L_{coh}$
may reach macroscopic
values. (For large variations of energy of the emitting particle,
$L_{coh}$ is given by an expression like (5). Paradoxically of this result 
is up to now being emphasized in review articles [6], 
although the results raises no doubts. However, in 1952 it was not at
once that one succeeded to convince even Landau and Pomeranchuk of the 
correctness of these arguments, of the presence of the formation zone 
increasing with energy, and so on [7]. Nevertheless, already in autumn 
1952 Ter-Mikayelian reported at Landau's seminar the details of his 
thesis, with complete mutual understanding and approval. It should be 
added only that the described effect was completely, in all theoretically 
developed details, checked experimentally in a crystalline medium, 
ten years later. At present it is used, in particular, to obtain 
quasimonochromatic and polarized ($\gamma$-quanta from electron
accelerators [8].

\indent
The importance of the arisen conception of the length of formation 
was at once estimated by Landau and Pomeranchuk, and they (and not only
they) began to think to further theoretically develop this phenomenon.

\vspace{3mm}
~~~~{\bf Influence of multiple scattering on the bremsstrahlung in 
amorphous medium.}\\
\indent
First Pomeranchuk noticed to Ter-Mikayelian that if all what he said 
about the coherence length in the crystal is correct, then in an amorphous
medium as well, the traditional Bethe-Hilter formula for the
bremsstrahlung on a single atom shuld have been changed, due to the
absorption on a distance termed the radiation length, at $L_{rad}\le
L_{coh}$. This statement raised no objections of either Ter-Mikayelian or 
Landau who advised to evaluate this effect. Soon, however, after examining
the problem, Landau came to the conclusion that the influence of multiple
scattering will take place rather than the influence of absorption by the
emitting particle. Rather soon Landau and Pomeranchuk evaluated this
effect and acquainted Ter-Mikayelian with the manuscript of their joint
article asking to tell his remarks. A discussion of this work took place,
and the article was approved. It happened so that Landau and Pomeranchuk,
starting with the formula (5), and explaining, that, in accordance with
the formula (6), the time tcoh "does very sharply increase with the energy
and, as a cosequence, those distances between electron and nucleus play a 
role which significantly exceed atomic sizes", did not, apparently, by a
misunderstanding only, refer to the work of Ter-Mikayelian. But he would
not think (felt shy?) to tell them that it had to be done [9]. This lead
to such a moving of events that this story seems to be not only quite
appropriate here but also instructive from the point of view of the 
scientific ethics.

\indent
In that time conditions of isolation of Soviet Science a 
publication of our works in foreign languages was strongly prohibited
as "cringing to abroad". Nevertheless, Landau's name in a published,
even in Russian, article, attracted the attention of the famous 
American physicist Dyson. Having, naturally, not known about the 
Ter-Mikayelian's work Dyson suspected that an interesting effect 
should exist in a crystal, and published (in coauthorship with G.
Ueberall) a paper in an american journal presenting a result 
coinciding with that of Ter-Mikayelian (and refered, of course, 
not to him but to Landau). Learning this Landau sent urgently the 
reprints of Ter-Mikayelian's works to Dyson, USA, showing that the
work he published had already been done here. In a reply letter 
Dyson appraised highly the works of Ter-Mikayelian and recognized 
that he with Ueberall obtained the same physical results using, 
however, another calculation technique. Landau there and then 
acquinted the participants of the next seminar with the Dyson's letter.

\indent
Being elegant and clear physically the work of Landau and 
Pomeranchuk needed some mathematical improvements. This has 
been done by A.B. Migdal [10] who used a fine and original 
technique to complete the Landau-Pomeranchuk theory, from 
quantitative point of view, to a logically closed form, and 
obtained an expression for the bremsstrahlung in amorphous 
medium with allowance for the influence of multiple scattering.
This expression used sometimes to be termed the Landau-Pomeranchuk-Migdal 
formula. Our physicists used this formula frequently to calculate the 
development of broad electromagnetic showers of cosmic rays. It is just 
this formula that was recently confirmed experimentally in Stanford.

\vspace{3mm}
{\bf Longitudinal density effect.}\\
\indent
With these works the investigations of pecuiliarities of the radiation
of ultrafast particles did not stop. Ter-Mikayelian generalized very 
soon the work of Landau-Pomeranchuk in he sense that he took into 
account the role of the dielectric polarization of the amorphous 
medium [11]. As it turned out, this polarizations affects the 
radiation of "soft" quanta with the energy of the order of or less than\\ 
\rightline
{{\large$\hbar\omega_{crit}=\hbar\omega_{0}/\sqrt{1-v^{2}/c^{2}}~.$}
~~~~~~~~~~~~~~~~~~~~~~~~~~~~~~~~~~~~~~~~~~~~~~~~~~~~~~~~~(7)}\\	
Here $\omega^{2}_{0}=4\pi NZe^{2}/m$ 
is the squared plasma frequency, $N$ the number of atoms per $cm^{3}$, 
$m$ and $e$ the electronic mass and charge, $Z$ the number of electrons in 
the atom. The estimation of influence of the medium polarization on 
the formation length can readily be obtained from the above expressions 
by taking into account that in a medium we have actually 
$k=\frac{\omega}{c}\sqrt{\varepsilon}$, with $\varepsilon$ being 
the dielectric constant of the medium. For the frequencies considerably 
exceeding the atomic ones:\\ 
\indent
$~~~~~~~~~~~~~~\varepsilon=1-\omega^{2}_{0}/2\omega^{2}$~.\\
Substituting correspondingly k in the expression (2), it is easy to see 
that the formula for the coherence length takes the form\\
\indent
{~~~~~~~~~~~~~~\large $L_{coh}=\frac{c}{\omega}\left [1/\left(1-\frac{v}{c}+
\frac{\omega^{2}_{0}}{2\omega^{2}}\right)\right]$}~.\\
This length is now "cut" for the photons at frequencies 
$\omega\le\omega_{crit}$. In 
this case, the increase in the energy of the emitting particle 
($v\rightarrow c$)
results in that $L_{coh}$ remains constant at a given frequency 
$\omega$, which 
leads to an essential modification of either the Bethe-Hitler or the
Landau-Pomeranchuk formula in the region of very soft quanta [12]. 

\indent
This density effect in the bremsstrahlung is in way similar 
to the density effect in ionization losses discovered by E. 
Fermi. The difference is following: in the second case it is 
the effective impact parameters (i.e. the distances in the 
direction perpendicular to paths of particles) that are "cut",
while in the first case it is the longitudinal distances along
the path of the emitting particle. In this connection the 
Stanford physicists term this phenomenon the longitudinal 
density effect with referring to the work [13] of Ter-Mikayelian.
Unfortunately, in the experiment performed in Stanford University, 
the intensity of emitted photons was measured in dependence 
on their energy only in the region from 5 to 500 MeV. 
Since the electron energy was 25 GeV the frequencies of those 
photons exceeded, and the longitudinal density effect could 
not still be displayed to the full extent. It would be interesting 
to conduct corresponding experiments (even at considerably lower 
energies of the emitting particle) for the emission spectrum of 
photons at frequencies of the order of or less than (crit. 
In principle, this method could be employed for measurements
of fast particle energies which is important for the 
experimental physics of ultrahigh energies.

\vspace{3mm}
{\bf Application to hadronic processes.}\\
\indent
We have already mentioned that the increase of the formation 
zone with the energy of relativistic particles, revealed by
Ter-Mikayelian, was applied also to high-energy hadronic processes. 
Here three consequent effects can be mentioned. The first, so to say, 
preliminary, is not of particular importance and has not been checked 
experimentally. It is valuable mainly from the methodological point of
view. With use of this effect it turned to be possible to calculate 
the emission of photons by a charged pion the plane wave of which is
incident no required to know the details of interaction between the
pion and nucleus, it is sufficient that such a nucleus cuts a round 
hole in the plane wave. Then a diffraction of pions occurs. And, as 
at small diffraction angles the length of zone of formation of emitted 
photon is very long, all needed integration can be made outside the 
nucleus and is performed without a detailed knowledge of the laws of 
interaction between the passing pion and nucleus [14]. However, the 
further attack in the same direction lead to much more important result.

\indent
A statement was made that a diffracted pion (like any hadron) 
can dissociate into other hadrons [15]. For example, 
a diffracted nucleon can emit a pion. Of course, in this case
the probability of such a diffraction dissociation, i.e. of the
process of pion emission by a diffracted hadron, can be calculated
only by the perturbative theory giving merely a rough estimation
of the cross-section of this process. But the cross-section is 
again determined by the integration over a large, increasing
with the energy region outside the hadron or nucleus target. 
This gives the process some features, which allow distinguishing 
it among other hadron generation processes. Prediction of the 
diffraction dissociation of hadrons raised doubts for a long 
time, but already in sixties it was confirmed experimentally,
and is now of great importance in high-energy hadron physics.
It was very concretely described in "Regestic" as an exchange of pomeron, 
a quasiparticle with zero quantum numbers [16].

\indent
However, in that "preRegestic" age many unclear question arose concerning 
the nature of the effect which, as we saw, seemed to occur completely 
outside the nucleus-target. It was questioned: "Where enters the 
interaction with the nucleus?" Once Pomeranchuk answered with irritation: 
"Well, you can hold that a chaste conception occurred".

\indent
In order to clear the mechanism, a special work has been done concerning
a similar possible effect, which is much more illustrative and calculable,
that is the effect of diffraction splitting (dissociation) of a deuton [17].
It was then developed to a new direction-diffraction splitting of nuclei.

\indent
Introduction of the concept of coherence length, or formation zone, its use
in various physical phenomena essentially changed our ideas about the 
radiation processes occurring at high energies. These are, we remind, 
in the first place, three effects that are under discussion here and 
confirmed experimentally: bremsstrahlung of photons in crystal, 
diffraction dissociation of hadrons and the Landau-Pomeranchuk 
effect in amorphous media. It should be noted that related processes 
have been considered earlier as well, particularly, when V.L. Ginsburg 
and I.M. Frank predicted the existence of transition radiation. The use
of concept of coherence length increasing with energy of emitting particle,
in consideration of this phenomenon permitted to enrich its theoretical
description, extend it considerably into the ultrahigh energy region,
 and then to create new detectors of relativistic particles [18].

\indent
All these works were done in a new years at the time when our country
separated from the world science by an "iron curtain". When this curtain
raised, the journal "Nuovo Cimento" ordered our scientists a number of
reviews of soviet investigations on various problems. It was found that
very much was done originally. As to the above-mentioned questions,
reviews on these topics [19] contained already more than a dozen original; 
publications which resulted actually from the work of Ter-Mikayelian. 

\vspace{3mm}
{\bf Footnotes}\\
1. See,, e.g., CERN Courier, 1994, v. 34, N 1, p. 12-13.\\
2. L.D. Landau , I.Ya. Pomeranchuk , Dokl. AN SSSR 92, 735 (1953).\\
3. M.L. Ter-Mikayelian , JETP 25, 289 (1953);  25, 296 (1953).\\
4. I had a pleasure to be his supervisor.\\
5. See footnote [3].\\
6. A.I. Akhiezer , N.F. Shulga , Uspekhi Fiz. Nauk 137, 561 (1982).\\
7. A colourful discussion with them on this occasion I have described 
in my memoirs. See: E.L. Feinberg  "Landau et al", Reminescence 
of  L.D. Landau, Moscow, 1988, p. 253.\\
8.For details see: M.L. Ter-Mikayelian , "High Energy Electromagnetic 
Processes in Medium", N.Y., 1972.\\
9. Later this misunderstanding was corrested . See: V.B. Berestetskii , 
E.M. Lifshits ,  L.P. Pitaevskii, Quantum Electrodynamics, Moscow, 1980, 
p. 452.\\
10. A.B. Migdal , Dokl. AN SSSR 54, (1954); 105, 77 (1955).\\
11. M.L. Ter-Mikayelian , Dokl. AN SSSR  94, 1033 (1954).\\
12. By treating this effect the same technique was used as in the work 
of Landau and Pomeranchuk. Migdal in his final publication of 1955 
(see footnote [10]) introduced also the corresponding changes into
his expressions.\\
13. See  Landau , I.Ya. Pomeranchuk , JETP 24, 505 (1953).\\
15. Pomeranchuk I.Ya, Feinberg E.L., Dokl. AN SSSR 93, 439 (1953).\\
16. See: P. Lanshof , Pomeron, Priroda, 1994, N 12, p. 17-25.\\
17. This phenomenon was predicted independently and practically at the 
same time by different authors. See: A.I. Akhiezer , A.G. Sitenko , 
JETP 32, 794 (1957); E.L. Feinberg, JETP 29 115 (1955); R. Clauber , Phys.
Rev., 88, 30 (19550.\\
18. See footnote [8]. For the relation between the transition radiation 
and the bremsstrahlung of ultrasort particles, see: M.L. Ter-Mikayelian, 
"Radiation of Particles in Periodic Media", Priroda, N 12, p. 68-73.\\
19. E.L. Feinberg , I.Ya Pomeranchuk , Nouvo Cimento, Supplemento, 
111 652 (1956); E.L. Feinberg , Usp. Fiz. Nauk, 58 193 (1956).

\newpage
\centerline{\bf{Figure captions}}
\indent
Fig.1.	DXR spectrum at$\Theta =42.9$ the 6.8 MeV electron beam direction.

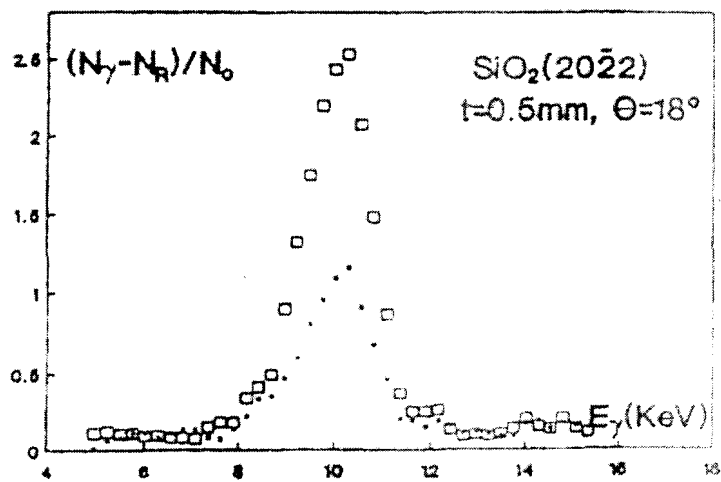
Fig.2. The spectra of electrons emission in quartz crystal under acoustic 
waves ($\Box$) and unaffected ($\bullet$).

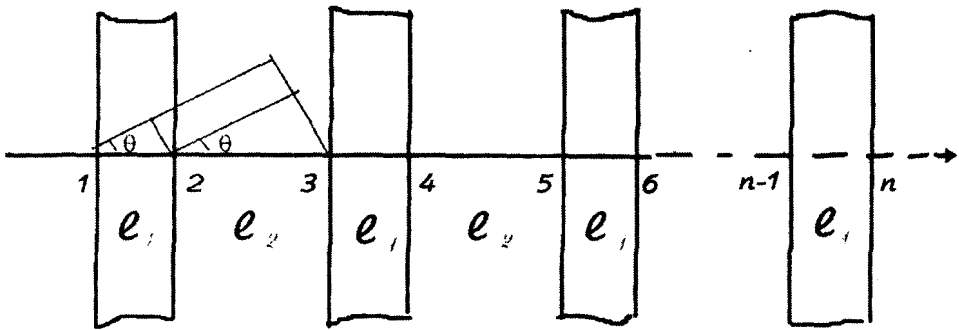
Fig.3. Particle passage through a stack of foils.

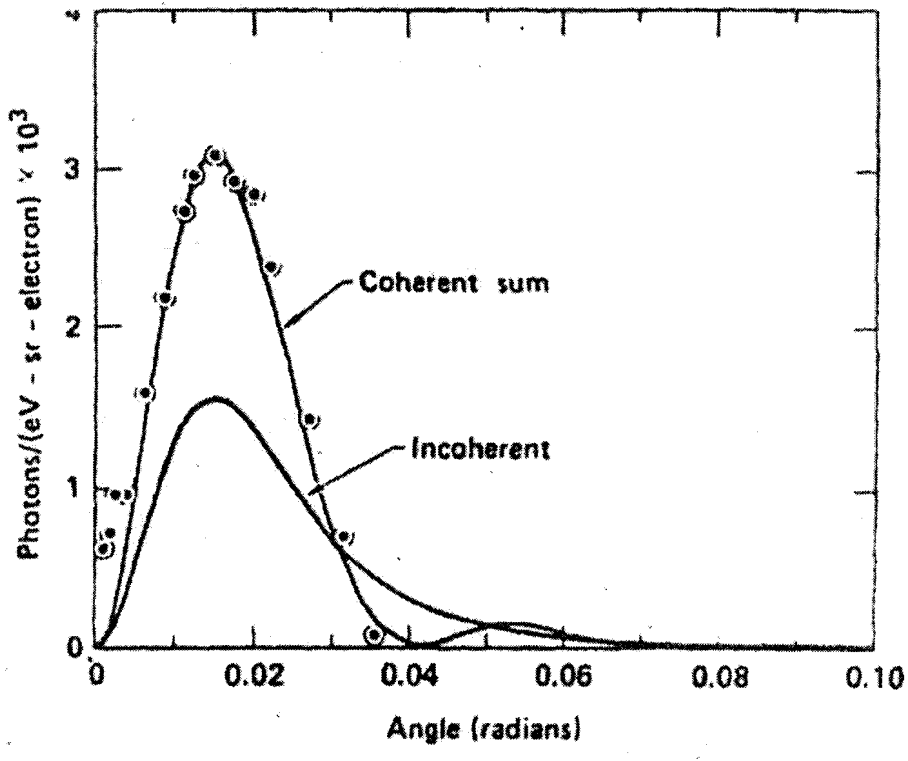
Fig.4. TR angular distribution in a plate. Thickness of beryllium foil was 
1 $\mu m$. Experiment demonstrates interference effect in a plate.

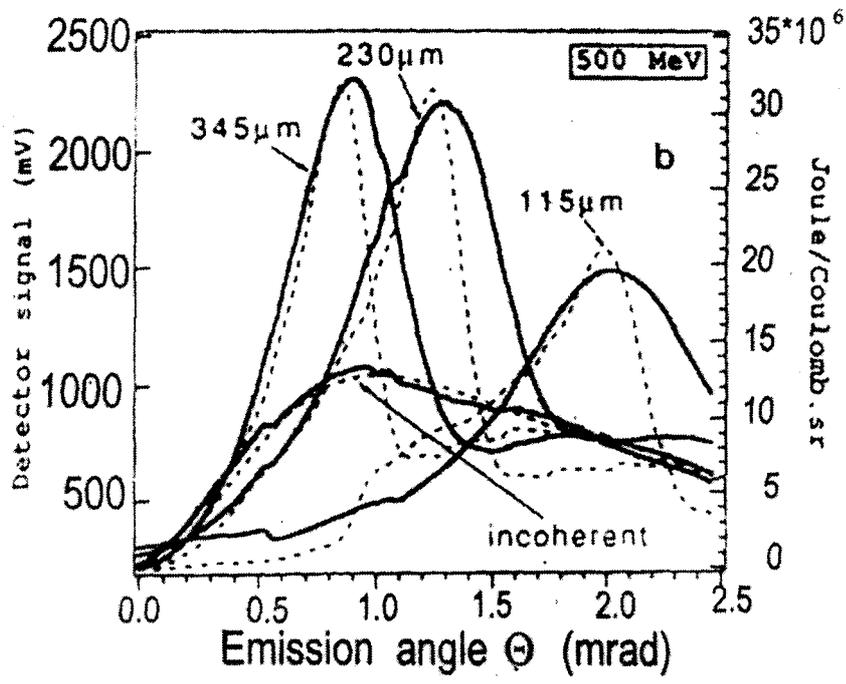
Fig.5. Integrated from 1 to 10 keV TR and RTR angular distribution from TR 
(incoherent) ($l_{2}$=1.5 mm) and RTR (coherent) ($l_{2}$=115, 230 and 345 
$\mu m$) stacks
 of 8 myler foils ($l_{1}=3.8 \mu m$).

FIg.6. i) The measured and calculated peak angle for TR (incoherent) and RTR (coherent); 
ii) The measured and iii) calculated spatial distribution for the (a) RTR,
 coherent and (b) TR, incoherent in myler stack.
 
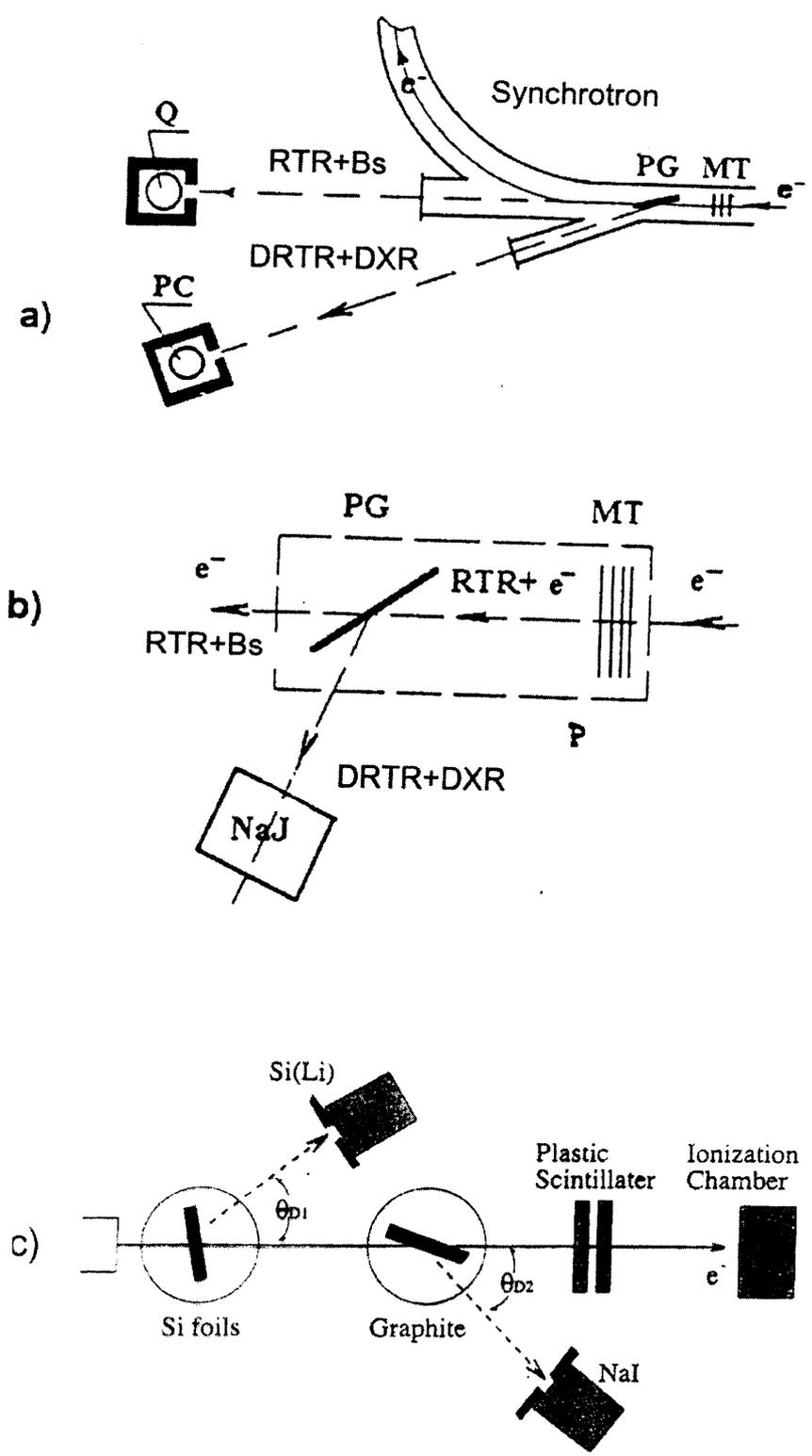
Fig.7. Experimental setups (Japan-Russian joint project).  

\begin{thebibliography}{99}
\bibitem{AR} M.L. Ter-Mikaelyan, "Vliyanie sredi na elektromagnitnie 
processi pri visokich energiyach", Acad. Nauk Armenii,Yerevan, 1969; 
"High Energy Electromagnetic Prosesses in Condenced Medium" John-Wiley 
$\&$ Sons, 1972. 
\bibitem{CA}  I.M. Dremin, Mod. Phys. Lett. A.13, N34, 2789 (1998); 
B. Kaempfer, Pavlenko, hep-ph/9906248 4 Jun 1999.
\bibitem{AD}  NIM, B145, Beam Interaction with Materials$\&$Atoms, 
(Ed.Andersen and L.E.Rehn), (Amsterdam-Tokyo, Elsevier, 1998). 
\bibitem{BE} Intr. Workshop on Radiation Physiks with Relativistic
electrons, (Ed.H. Backe, W. Lauth, T. Walcher), (Tabarz), 
Germany, 1998, Abstract
\bibitem{NA} E L. Feinberg,  Priroda 1, 30 (1994).   
\bibitem{DA}  C.A. Vorobyov , B.N. Kalinin, C.D. Pak, A.P. Potylitsin,
 JETP Lett. 41, 1 (1985).
\bibitem{RA} Yu.N. Adishchev, V.G. Baryshevski, C.A. Vorobyov et al., 
JETP Lett. 41, 361 (1985); Yu.N. Adishchev, S.A. Vorobiev, B.N. Kalinin, 
S. Pak, A.P. Potylitsin, JETP 90, 829 (1986).
\bibitem{DE} R.O. Avakyan, A.E. Avitisyan, Yu.N. Adishchev et al., JETP
 Lett. 45, 313 (1987).
\bibitem{CB} A.V. Shchagin, V.I. Pristupa, N.A. Khizhnyak,  Phys. Lett. A 148, 485 
(1990); NIM B99, 277 (1995); A.V. Shchagin, N.A. Khizhnyak, NIM B119, 115 (1996).
\bibitem{CE} V. Afanasenko, V.G. Baryshevski  et al., Phys. Lett. A 170, 315 (1992).
\bibitem{CF} S. Asano, I. Endo, M. Harada et al., Phys. Rev. Lett. 70, 3247 (1993).
\bibitem{CI} R.B. Fiorito, D.W. Rule et al., Phys. Rev. Lett. 71, 704 (1993).
\bibitem{CJ} A.V. Shchagin, V.I. Pristupa, N.A. Khizhnyak, Proc. of RREPS, 
1993, pp 62-75, Tomsk, Russia. NIM B99, 277 (1993); NIM, B119, 115 (1996).
\bibitem{CO} H. Nitta, Phys. Rev. B45, 4 7645 (1992); Phys. Lett. 
A 158, 270 (1991); NIM B115, 401 (1996).
\bibitem{CL}  V.G. Baryshevski, "Channeling, radiation and reaction in crystals in 
high energies", Minsk University (1982).
\bibitem{CK} G.M. Garibian, C. Yang, "Transition X Radiation", Yerevan, 
Acad. of Sience (1983).
\bibitem{CR} J. Freudenberger, M. Galemann, H. Genz et al., NIM B115, 408 (1996).
\bibitem{CG} J. Freudenberger, H. Genz, V.V. Morokhovski, Appl. Phys. Lett. 70, 2, (1997).
\bibitem{EA} K.H. Brenzinger, B. Limburg, H. Backe et al., Phys. Rev. Lett. 79, 2462 
(1997); Zeit. Fuer Phys. A 358, 107 (1997).
\bibitem{EB} Th. Doerk, H. Backe, N. Clawiter et al, Intern. Workshop, 
Tabarz, Ger., June 9-12, Abst.
\bibitem{EC} V.V. Morokhovski, K.N. Schmidt, G. Buschhorn et al., Phys.Rev. Lett. 
79, 4389 (1997); V.V. Morokhovski, Dissertation D17, Technische Universitaet Darmstadt 1998.
\bibitem{ED} K.H. Schmidt, G. Buschhorn, R. Kottaus et al., NIM B145, 8 (1998). 
V.V. Morokhovskii, J. Freudenberger, H. Genz, et al., NIM B145, 14 (1998).
\bibitem{EF} R. Kottaus, G. Buschhorn, M. Rzepka et al., Conference, San-Diego, July, 
1998, S.P.I.E., vol. 3443, 105 (1998).
\bibitem{EI} A.V. Shchagin, Phys. Lett., A 247, 27 (1998).
\bibitem{EG} M.L. Ter-Mikaelyan, Dokladi Acad. of Sciences of USSR 94, 1033 (1954).
\bibitem{EH} Yu.N. Adishchev, V.A. Verzilov, S.A. Vorobyev et al., JETP Lett., 48, 311 
(1988); NIM B44, 130 (1989).
\bibitem{EK} I. Endo, M. Harada, T. Kabayashi et al., Phys. Rev. E 51, 6305 (1995).
\bibitem{EL} R.B. Fiorito, D.W. Rule, M.A. Piestrup et al., Phys. Rev. E 51, 2759 (1995).
\bibitem{EQ} J. Freudenberger, Phys. Rev. Lett. 74, 2487 (1995).
\bibitem{FA} I.D. Feranchuk, A.V. Ivashin, J. Phys Paris 46, 1981 (1985).
\bibitem{FB} K.Yu. Amosov, B.N. Kalinin, A.P. Potylitsin et al., Phys. 
Rev. E 47, 2207 (1993).
\bibitem{FC} R.O. Avakian, A.E. Avetissian, H.S. Kizoghian et al., Radiation Effects 
and Defects in Solids 117, 17 (1991). H.A. Aslanyan, A.H. Mkrtchyan, 
Inter. Workshop, Tabarz, Germany, June, 1998, Abstract; Phys. Lett. A152 297 (1991).
\bibitem{FD} R.B. Fiorito, D.W. Rule et al., NIM B79, 758 (1993).
\bibitem{FE} G.M. Garibian, JETP 37, 527 (1959).
\bibitem{FG} K.A. Barsukov, JETP 37, 1106 (1959).
\bibitem{FH} M.L. Ter-Mikaelyan, JETP 25, 289 (1953); 25, 296 (1953).
\bibitem{FI} I.M. Frank, UFN 75, 231 (1961).
\bibitem{FL} M.L. Ter-Mikaelyan, Doklady Ak.Nauk USSR 134, 318 (1960).
\bibitem{FK} M.L. Ter-Mikaelyan,  Nuclear Physics 24, 43 (1961); JETP 
Letters 8, 100 (1968).
\bibitem{FN} F.R. Arutyunyan, K.A. Ispiryan,  Yadernaya Fizika 1, 842 (1964).
\bibitem{FM} F.R. Arutyunyan, M.L. Ter-Mikaelyan, UFN 107, 325 (1972).
\bibitem{FQ} V.L. Ginzburg,  Priroda 8, 56 (1975); V.L. Ginzburg, 
V.N. Tsytovich, Phys. Report 49, 1 (1979).
\bibitem{FS} M.L. Cherry, G. Hartmann, Phys. Rev. D 10, 3594 (1974).
\bibitem{IA} X. Artru, G. Yodh, G. Mennessier, Phys. Rev. D 12, 1228 (1975).
\bibitem{IB} M. Deutschmann, W. Struczinski, C.W. Fabjan et al., NIM 180, 409 (1981).
\bibitem{IC} B. Dolgoshein, NIM A 326, 434 (1993).
\bibitem{ID} P.J. Ebert, M.J. Moran, B.A. Dahling et al., Phys. Rev. 
Lett. 54, 9, 893 (1985).
\bibitem{IE} M.A. Piestrup, J.O. Kephart, R.K. Park et al., Phys. Rev. 
A.32, 917 (1985).
\bibitem{IF} M.A. Piestrup, D.G. Boyers, C.I. Pinkus et al, Appl. Phys. 
Lett. 58, 23 2692 (1991); 59, 2 189 (1991).
\bibitem{IJ} M.J. Moran, B.A. Dahling, P.J. Ebert et al, Phys. Rev. 
Lett. 57, 1223 (1986).
\bibitem{IG} P. Goedtkindt, J.M. Salome, X. Artru et al., NIM B56-57, 1060 (1991).
\bibitem{IL} P. Goedtkindt, J.M. Salome, X. Artru et al., Microelec. 
Engin., 13, 327 (1991).
\bibitem{IN} M.A. Piestrup, D.G. Boyers, C.I. Pincus et al., Phys. Rev. 
A. 45, 2, 1183 (1992).
\bibitem{IM} T. Tanaka, T. Awata, A. Itoh et al., NIMB 93, 21 (1994).
\bibitem{IR} T. Awata, K. Yajima, T. Tanaka et al., Radiat. Phys. Chem. 
50, 3, 207 (1997).
\bibitem{IQ} T. Awata, K. Yajima, T. Tanaka et al., Application of Accelerator
in Research and Industry, edited by J.L. Duggan and I.I. Morgan, A.I.P., New-York, 1997.
\bibitem{LA} T. Awata, K. Yajima, Y. Koizumi et al., Beam Science 
and Technology 3, 10 (1998).
\bibitem{LB} S. Asano, I. Endo, M. Harada et al., Phys. Rev. Lett. 70, 3247 (1993).
\bibitem{LC} A.P. Potylitsyn, V.A. Verzilov, Phys. Lett. A 209, 380-384 (1995).
\bibitem{LD} M.Yu. Andreyashkin, V.V. Kaplin, A.P. Potylitsin et al, NIM B119, 
108 (1996); M.Yu. Andreyashkin, B.N. Zabaev, V.V. Kaplin et al., 
Pisma JETP 65, 8, 594 (1997).
\bibitem{LE} Y. Takashima, K. Aramitsu, I. Endo, NIM B145, 25 (1998).
\bibitem{LF} A.P. Potylitsyn, Physics Lett. A 238, 112 (1998).
\bibitem{LI} A.P. Potylitsyn, NIM B145, 60 (1998).
\bibitem{LN} V.G. Baryshevsky, NIM A122, 13 (1997).
\bibitem{LM} X. Artru, P. Rullhusen, NIM B145, 1 (1998).
\bibitem{LR} N. Nasonov, Phys. Lett. A 246, 148 (1998).
 
\end{thebibliography}
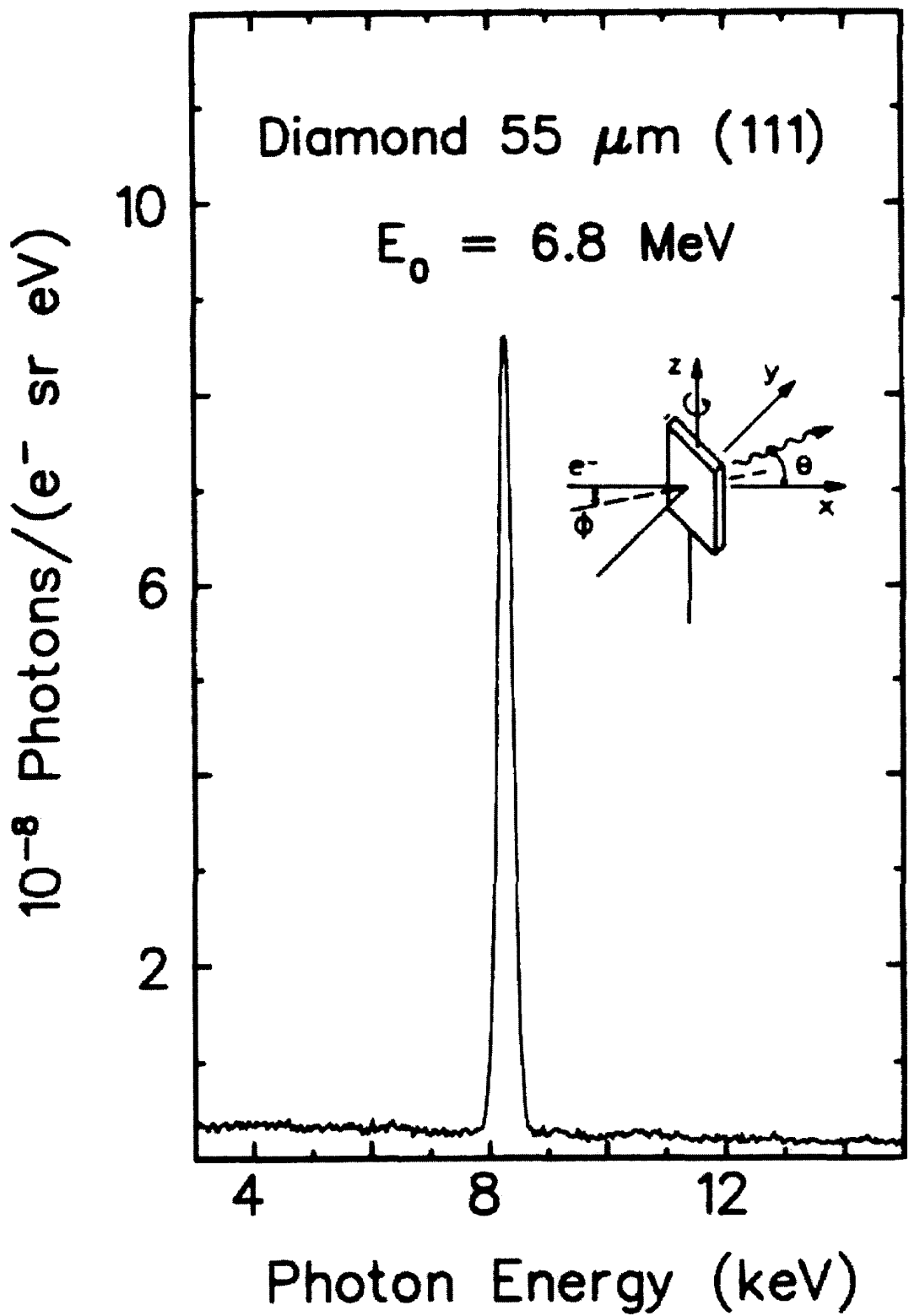
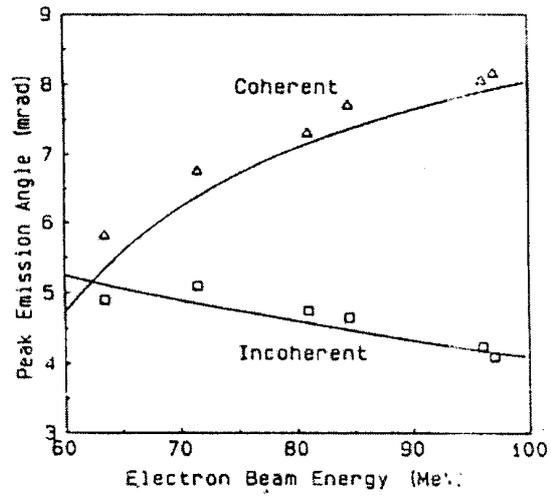
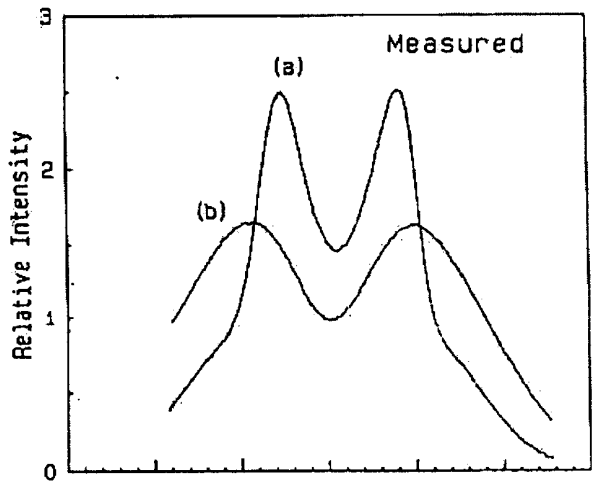
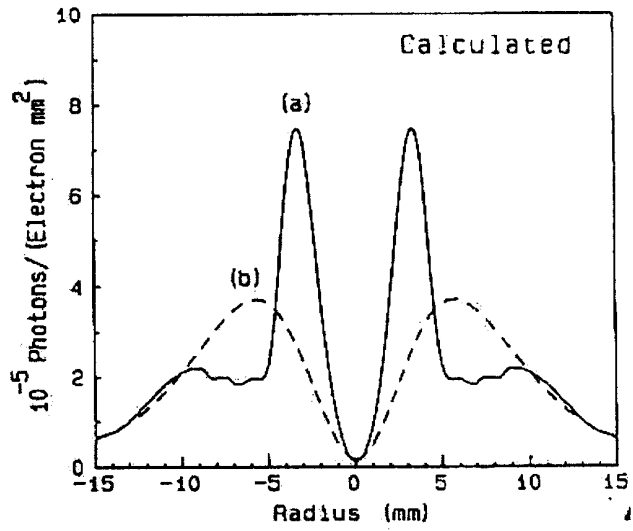
\end{document}